\documentclass[12pt,preprint]{aastex}

\newcommand{\gx}{GX\,339--4}
\newcommand{\HeII}{He\,{\sc ii}}
\newcommand{\CIII}{C\,{\sc iii}}
\newcommand{\NIII}{N\,{\sc iii}}
\slugcomment{Accepted for publication in Astrophysical Journal Letters}

\shorttitle{Dynamical Evidence for a Black Hole in GX\,339--4}
\shortauthors{Hynes et al.}

\begin{document}


\title{Dynamical Evidence for a Black Hole in GX\,339--4}


\author{R. I. Hynes\altaffilmark{1,4,7}, 
  D. Steeghs\altaffilmark{2,4},
  J. Casares\altaffilmark{3}, 
  P. A. Charles\altaffilmark{4}, 
  and
  K. O'Brien\altaffilmark{5,6}}

\altaffiltext{1}{The University of Texas at Austin,
  Astronomy Department, 1 University Station C1400, Austin, Texas
  78712, USA; rih@astro.as.utexas.edu}
\altaffiltext{2}{Harvard-Smithsonian Center for Astrophysics, 
60 Garden Street, MS-67, Cambridge, MA 02138, USA; 
dsteeghs@head-cfa.harvard.edu}
\altaffiltext{3}{Instituto de Astrof\'\i{}sica de Canarias, 38200 La Laguna,
Tenerife, Spain; jcv@ll.iac.es}
\altaffiltext{4}{Department of Physics and Astronomy, 
University of Southampton, Southampton, SO17 1BJ, UK; pac@astro.soton.ac.uk}
\altaffiltext{5}{European Southern Observatory, Casilla 19001,
 Santiago 19, Chile; kobrien@eso.org}
\altaffiltext{6}{School of Physics and Astronomy, University of St
  Andrews, St Andrews KY16 9SS, UK}
\altaffiltext{7}{Hubble Fellow}


\begin{abstract}
We present outburst spectroscopy of \gx\ which may reveal the motion
of its elusive companion star.  \NIII\ lines exhibit sharp emission
components moving over $\sim300$\,km\,s$^{-1}$ in a single night.  The
most plausible interpretation of these components is that they are
formed by irradiation of the companion star and the velocities
indicate its orbital motion.  We also detect motion of the wings of
the \HeII\ 4686\,\AA\ line and changes in its morphology.  No
previously proposed period is consistent with periodic behavior of all
of these measures.  However, consistent and sensible solutions are
obtained for periods around 1.7\,days.  For the best period,
1.7557\,days, we estimate a mass function of $5.8\pm0.5$\,M$_{\odot}$.
Even allowing for aliases, the 95\,\%\ confidence lower-limit on the
mass function is 2.0\,M$_{\odot}$.  \gx\ can therefore be added to the
list of dynamical black hole candidates.  This is supported by the
small motion in the wings of the \HeII\ line; if the compact object
velocity is not larger than the observed motion then the mass ratio is
$q \la 0.08$, similar to other systems harboring black holes.
Finally, we note that the sharp components are not always present, but
do seem to occur within a repeating phase range.  This appears to
migrate between our epochs of observation, and may indicate shielding
of the companion star by a variable accretion geometry such as a warp.
\end{abstract}


\keywords{accretion, accretion disks --- binaries: close --- X-rays:
  binaries --- stars:
  individual: V821~Ara}


\section{Introduction}

Most Galactic black holes are found in transient low-mass X-ray
binaries (LMXBs), where bright X-ray outbursts are interspersed with
long quiescent periods when the optical light is typically dominated
by the companion star.  Although one of the earliest proposed black
hole candidates \citep{Samimi:1979a}, \gx\ remains one of the most
inaccessible to direct measurement of system parameters due to a
higher level of X-ray activity; even when the X-rays are at their
faintest, the companion star remains undetectable
\citep{Shahbaz:2001a}.  Several orbital periods have been proposed
based on both photometric and spectroscopic observations (0.62\,days;
\citet{Callanan:1992a}; 0.7\,days; \citet{Cowley:2002a}), but even
this most basic of binary parameters remains in dispute.  Without
detecting the companion star in quiescence, this problem seems
insurmountable.  However an approach was suggested by the discovery by
\citet{Steeghs:2002a} of sharp \NIII\ emission components in the
bright persistent low-mass X-ray binary Sco X-1.  They were associated
with the irradiated companion star and yielded the first radial
velocity curve of the companion, and hence the first credible estimate
of the binary parameters.

When \gx\ entered a bright X-ray outburst in mid-2002, we sought to
apply this method.  Initial observations used the NTT and AAT;
follow-up VLT observations were also obtained.  The average {\it
RXTE}/ASM count rates\footnote{Based on quick-look results provided by
the ASM/RXTE team.} corresponded to 0.46\,Crab at the first epoch and
0.84\,Crab at the second.  Here we report the detection of the
companion star in the \NIII\ lines, a new orbital period estimate, and
the derived mass function of the compact object.  A future work will
perform a thorough analysis of other aspects of this rich dataset.

\section{Observations}

Medium dispersion spectroscopy was obtained with the ESO NTT on 2002
June 8--11 using EMMI.  Details of all our observations are given in
Table~\ref{ObsTable}.  On June 8 and 10 we operated the red-arm with
the MIT/LL detector and grating \#6, yielding a wavelength coverage of
4410--5160\,\AA.  In poorer conditions on June 9 we used the blue arm,
TK1034, and grating \#3 resulting in a lower resolution and wavelength
coverage 4500--4960\,\AA.  Data reduction was done in {\sc iraf} to
remove the bias level, flat-field, and optimally extract the spectra
\citep{Horne:1986a}.  Wavelength calibrations used contemporaneous
HeAr (red-arm) or ThAr (blue-arm) arc lamp observations.  The slit was
aligned to also include a nearby K star (star 2 of
\citet{Grindlay:1979a}).  This was rich in absorption lines and so
allowed a frame-by-frame check of the stability of the wavelength
calibration.

We also used the RGO Spectrograph attached to the AAT over 2002 June
6--11, with the R1200B grating yielding a wavelength coverage of
3500--5250\,\AA.  The images were de-biased and flat-fielded, and the
spectra subsequently extracted using conventional optimal extraction
techniques \citep{Horne:1986a}.  Wavelength calibrations were
interpolated between CuAr comparison lamp images, obtained every
20--30\,mins.

A final series of spectra was obtained using UVES at the VLT on 2002
August 9--15.  Six sets of five spectra were obtained in an identical
configuration; the red-arm was used with the EEV+MIT/LL mosaic and
cross-disperser \#3, yielding a wavelength coverage of
4100--6200\,\AA.  Pipeline optimal extractions were supplied and were
of good quality.  Some spectra exhibited a narrow dip within the Bowen
blend; these appear spurious and were masked out of our fits.
%
%
\section{Radial velocities and the orbital period}

\subsection{The sharp Bowen components}
\begin{figure}
\plotone{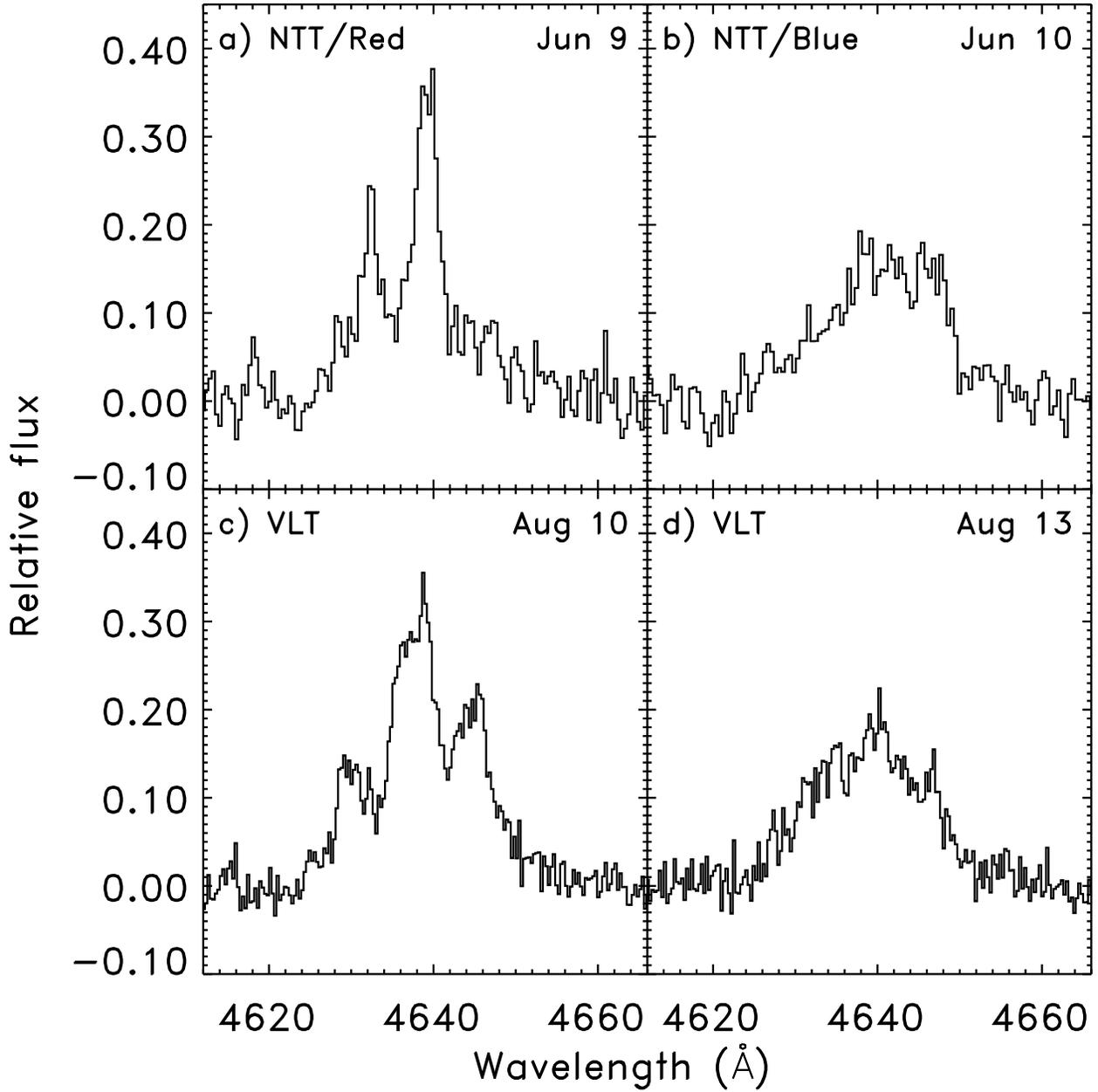}
\caption{Examples of Bowen blend profiles from four different
observations.  The left hand two panels show profiles with sharp
components; the right hand panels lack them.  
All spectra have been normalized and then
continuum subtracted, so fluxes are relative to the continuum level.}
\label{ProfileFig}
\end{figure}

The spectrum around 4640\,\AA\ is dominated by two or three emission
components (e.g.\ Fig.~\ref{ProfileFig}).  These are \NIII\ 4634 and
4641/42\,\AA, which are always present and \CIII\ 4647/50/51\,\AA,
which is sometimes as strong as the \NIII\ lines and sometimes nearly
absent.  This decomposition is the same as suggested for Her X-1
\citep{Still:1997a} and Sco X-1 \citep{Steeghs:2002a}.  We will refer
to these lines as the Bowen blend, although the \CIII\ emission is not
formed by Bowen fluorescence.  For a radial velocity analysis, we
selected only those spectra with sharp components.  The resulting
dataset comprised all the NTT red-arm spectra and 7 of the VLT
spectra.  No sharp components were detected in the AAT or NTT blue
spectra, neither were they present in the remainder of the VLT
spectra.  In our `sharp component' spectra, we measure average
component FWHM of 270\,km\,s$^{-1}$ at the June epoch and
350\,km\,s$^{-1}$ at the second, with no significant width modulation
within each epoch.

For each spectrum showing sharp components, we fit with a model
comprising two sets of Gaussian lines, corresponding to \NIII\
4634/41/42\,\AA\ and \CIII\ 4647/50/51\,\AA.  The relative ratios
within each species were fixed to the values given by
\citet{McClintock:1975a}.  We allowed the relative strengths of the
\NIII\ and \CIII\ groups to vary, and fitted for the \NIII\ and \CIII\
velocities.  The \CIII\ velocities were discarded as these lines were
often barely detectable, and the uncertain ratios of the unresolved
4647/50/51\,\AA\ components renders the effective wavelength of the
\CIII\ blend uncertain.  The \NIII\ velocities were always well
constrained, as the 4634\,\AA\ line is well separated from the others.
The most dramatic changes were in the NTT red-arm data, for which we
obtained a continuous night of data, about 9\,hrs.  During this period
we saw a monotonic trend spanning $\sim300$\,km\,s$^{-1}$.  Such
narrow components moving at high radial velocity can only readily be
explained as due to the orbital motion of the companion star.  The
measured components widths are larger than the expected rotational
broadening, in particular during our VLT observations. Outflows driven
by the X-ray irradiation may be one source of additional broadening
mechanisms.

\begin{figure}
\plotone{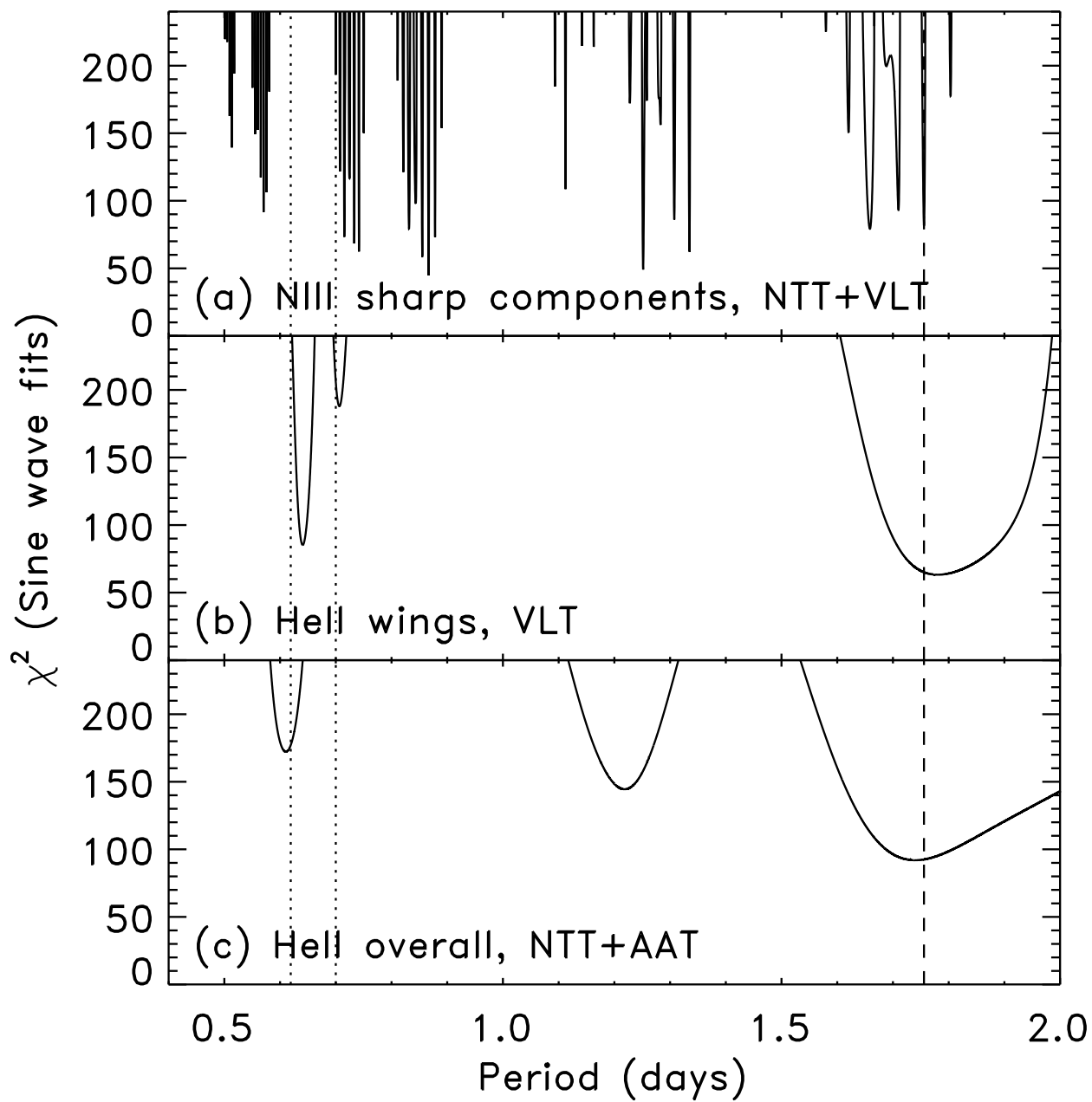}
\caption{$\chi^2$ of sinusoidal model fits as a function of orbital
period.  The systemic velocity, radial velocity semi-amplitude and
phasing are allowed to vary to obtain the best fit of each dataset for
a given orbital period.  The dashed line shows our preferred
1.7557\,day period; dotted lines indicate the 0.62 and 0.7\,day
periods of \citet{Callanan:1992a} and \citet{Cowley:2002a}
respectively.}
\label{PeriodFig}
\end{figure}

\begin{figure}
\plotone{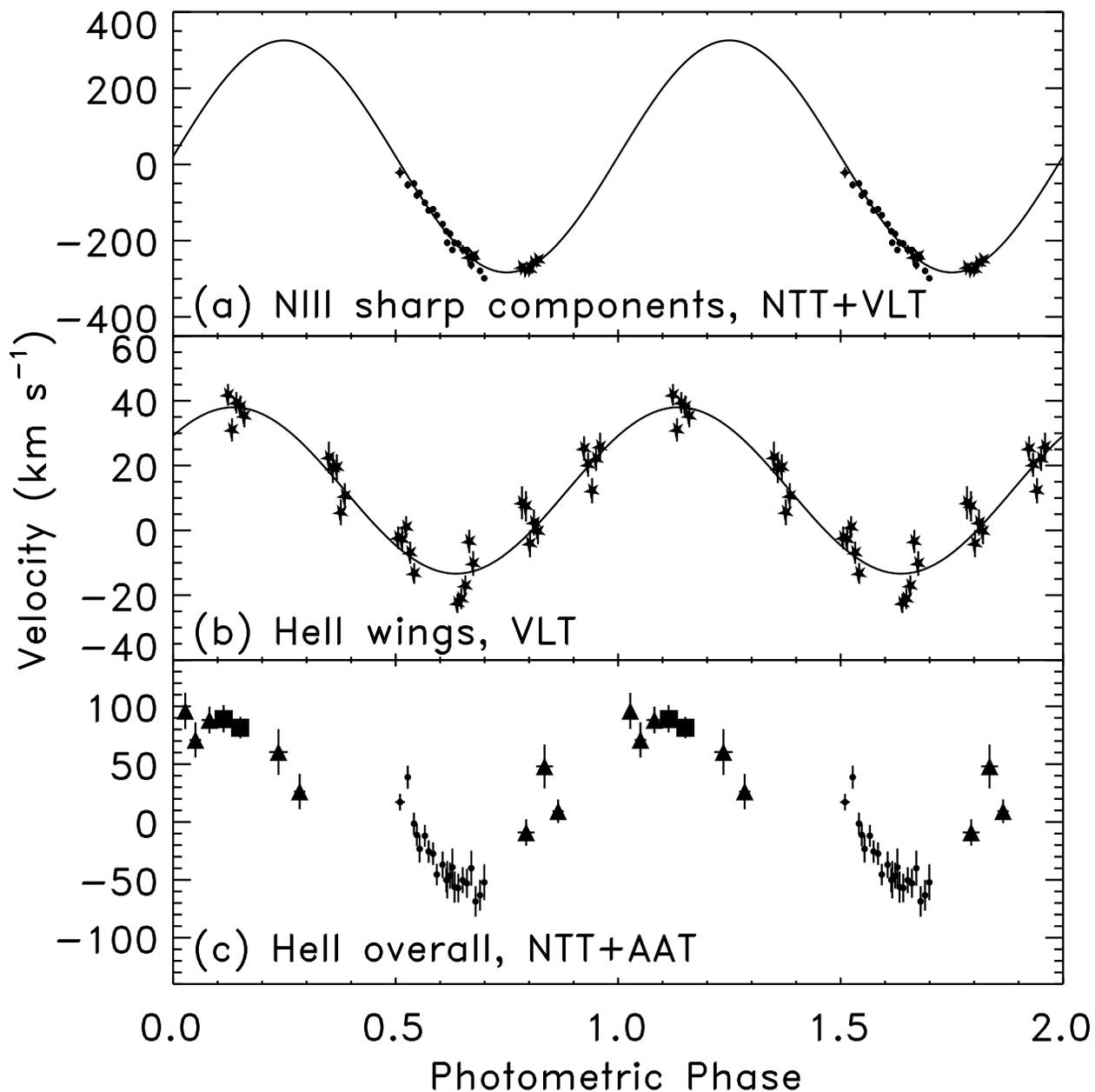}
\caption{Radial velocity curves corresponding to the data used in
Fig.~\ref{PeriodFig}, folded on our preferred 1.7557 day period.
Circles denote NTT red-arm data, squares NTT blue-arm, triangles AAT,
and stars VLT data.  Where error-bars are not apparent, they are
smaller than the points.  No fit is shown for the third panel as this
reflects changes in morphology more than actual motion and so the
radial velocity curve has no physical significance, other than as a
diagnostic of the period.  Note that (b) and (c) are derived from
completely independent sets of data.}
\label{RVCurveFig}
\end{figure}

We combined all of the velocities and fitted sinusoidal radial
velocity curves as a function of orbital period.  We used a grid of
trial periods from 0.1--2.5\,days, with a separation of
$10^{-4}$\,days, enough to resolve individual minima.  We plot the
minimum $\chi^2$ as a function of period in Fig.~\ref{PeriodFig}a.
There are clearly many possible solutions, corresponding to periods
around 0.7, 0.85, 1.2, or 1.7\,days, but the 0.62\,day period of
\citet{Callanan:1992a} is ruled out.  As we will argue, the preferred
period is 1.7557\,days, and the \NIII\ velocities are folded on this
period in Fig.~\ref{RVCurveFig}a.

\subsection{\HeII\ wings}

To further constrain the orbital period, we applied the double
Gaussian method of \cite{Schneider:1980a} and \cite{Shafter:1983a} to
trace the motion of the \HeII\ line wings in the VLT spectra.  These
may follow the orbit of the compact object, but even if they are
contaminated then any motion should still be on the orbital period.
We use all of the VLT spectra for this analysis.  As for the \NIII\
velocities, we fit a sinusoidal radial velocity curve as a function of
orbital period and derive the periodogram shown in
Fig.~\ref{PeriodFig}b.  The only periods common to both the \NIII\
data and the \HeII\ line wings are around 0.7 and 1.7\,days.  We
initially favored the former, as this value had previously been
suggested by \citet{Cowley:2002a}.  A significant difficulty with this
period, however, is that nearly all the \NIII\ velocities measured
correspond to blue-shifts.  For an 0.7\,day period, almost the full
range of radial velocities should have been sampled and so the
systemic velocity of the companion star would have to be extremely
large, $\sim-150$\,km\,s$^{-1}$.  The systemic velocity inferred from
the \HeII\ wings, however, is small, only 10\,km\,s$^{-1}$.  This
inconsistency led us to consider whether the longer period might
actually be the correct one as this results in consistent and small
systemic velocities.

\subsection{Line morphology changes}

An outstanding question is why only some \NIII\ profiles show sharp
components, all corresponding to blue-shifts.  One might naturally
expect this to reflect orbital morphology changes, but this is not the
case for the 0.7\,day period; in this case, spectra showing sharp
components overlap in phase with those which do not.  On the 1.7\,day
period, however, the two types of profiles are cleanly segregated and
do not overlap; this supports the idea that we are seeing orbital
motion of a companion star which is only visible, or only illuminated,
at certain orbital phases.

A related problem is that in the long NTT run, the \HeII\ line always
showed a blue-shifted peak.  As for \NIII, this is hard to explain
with an 0.7\,day period as other spectra at the same implied orbital
phase do not show this asymmetry.  We can quantify it by fitting a
single broad Gaussian profile to \HeII, and this line is strong enough
that we can use {\em all} of our first epoch data.  The derived radial
velocity curve will not have much physical significance, but this does
provide a quantitative way to assess which trial periods yield
repeatable changes in the line symmetry.  The periodogram, derived as
above, is shown in Fig.~\ref{PeriodFig}c.  As for the appearance of
sharp Bowen components, the \HeII\ symmetry changes are modulated on
the 1.7\,day period, but not clearly on the 0.7\,day one.

\subsection{The derived orbital period}

We conclude that only a period close to 1.7\,days is consistent with
the \NIII\ velocities and appearance of sharp components; the \HeII\
wing velocities; and the changes in the \HeII\ morphology.  This
period has the advantage, unlike the others, that a consistent and
sensible (i.e.\ close to zero) systemic velocity is derived for both
\NIII\ lines (which we believe trace the companion star velocity) and
\HeII\ wings.  Several aliases (from the most precise Bowen fit) are
consistent with the \HeII\ data.  The best overall fit is for $P_{\rm
orb} = (1.7557\pm0.0004)$\,days, but $1.7104\pm0.0006$\,days or
$1.6584\pm0.0017$\,days are also possible.

%
\section{The system parameters}
Having identified the best-fitting orbital period, we can now fit for
orbital velocities.  The most straightforward is the \NIII\
modulation, which must originate from the companion star.  We show the
fit to the data in Fig.~\ref{RVCurveFig}a.  The limited phase coverage
obviously compromises the accuracy of our results, and we find there
is a range of solutions with $K_2$ and $\gamma$ positively correlated.
For the 1.7557\,day period, the best fit is $K_2 = (317 \pm
10)$\,km\,s$^{-1}$, where the error is a 1\,$\sigma$ confidence region
with $\gamma$ left unconstrained.  This best fit corresponds to
$\gamma\sim 30$\,km\,s$^{-1}$, measured with respect to the local
standard of rest.  For the 1.7104\,day period $K_2 = (289 \pm
10)$\,km\,s$^{-1}$ and for 1.6584\,days, $K_2 = (241 \pm
8)$\,km\,s$^{-1}$.  A more conservative constraint is to consider all
three aliases as equally valid.  We then derive a 95\,\%\ confidence
lower limit on $K_2$ of 226\,km\,s$^{-1}$, corresponding to the
1.6584\,day period.  These values are all strictly lower limits for
$K_2$, as \NIII\ lines will be formed on the inner, irradiated face of
the companion star.  Thus they do not trace the motion of its center
of mass, but instead have a somewhat {\em lower} velocity.

From these combinations of $P_{\rm orb}$ and $K_2$, we then derive
corresponding mass functions.  Our best period corresponds to $f(M) =
(5.8\pm0.5)$\,M$_{\odot}$, whereas the 95\,\%\ lower limit is
2.0\,M$_{\odot}$.  The best fit mass function significantly exceeds
the canonical maximum mass of a neutron star (3.0\,M$_{\odot}$), and
even allowing for the shorter period alias, a black hole seems
likely, especially as $K_2$ has probably been underestimated.  

In principle, we could hope that the wings of the \HeII\ line will
follow the compact object.  Hence we might derive $K_1$ and the mass
ratio.  In this case, however, the \HeII\ velocities should be
anti-phased with the \NIII\ ones, which is clearly not the case.
Rather they are almost in phase with it, as is the overall movement of
the \HeII\ line.  This indicates that even the radial velocity curve
of the wings is significantly contaminated by \HeII\ emission from the
companion and/or the stream-impact point; the phase offset is in the
direction expected for emission from the stream-impact point, as is
commonly seen in other LMXBs and cataclysmic variables.  At most we
can then hope that the low amplitude of the observed modulation rules
out a large $K_1$.  From the VLT data we measure a semi-amplitude of
$25.7\pm1.4$\,km\,s$^{-1}$.  If we then assume $K_1$ is not larger
than this, then we expect a mass ratio $q\la0.08$, similar to other
black hole candidates.  This estimate is not secure, however, as one
could imagine a situation where contamination of the radial velocity
curve could cancel out a larger $K_1$ reducing the overall modulation
to close to zero.

%
%
\section{Discussion}

Our estimate of the mass function is surprisingly large.  It has been
widely assumed that the inclination of \gx\ is low; hence we would
expect a lower $K_2$ and a small mass function.  The most convincing
argument, by \citet{Wu:2001a}, is based on the small peak separation
of double-peaked lines.  This is only weakly sensitive to our revised
orbital period and is hard to discount.  Interpretations of peak
separations are not straightforward, however, and when the parameters
are known a priori, apparent sub-Keplerian motions are often inferred
(e.g.\ \cite{Marsh:1998a}).  Hence the peak separation is not a
reliable diagnostic of the system parameters.  Nonetheless, the
inconsistency of a small peak separation and our moderately large mass
function is the most problematic feature of our interpretation.

It is puzzling, and frustrating, that sharp \NIII\ components are not
present all the time, instead appearing confined to a limited phase
range.  This is why our radial velocity curve is incomplete, despite
good orbital phase coverage.  The phase range where they are seen is
different at the two epochs; in our last VLT observation we see them
emerge during the sequence, at around phase 0.65, whereas the NTT
observations revealed them earlier than this.  One possible
interpretation involves variable shielding of the companion star by a
warped disk.  The smaller mass ratios of black hole systems compared
to those containing neutron stars will reduce the angular size of the
companion Roche lobe as seen by the X-ray source, so shielding by the
disk will be more significant in the black hole case.  If the disk is
warped then each hemisphere of the companion may only be illuminated
during part of the orbit, as the companion rotates around the nearly
stationary warp pattern.  As the warp slowly precesses, the
illuminated phase range will migrate as we seem to observe.  Of
course, the X-ray brightness was also different at the two epochs and
other changes in the irradiation geometry could also be responsible
for the difference.
%
%
\section{Conclusions}
We have shown that optical spectroscopy of the Bowen blend and \HeII\
are consistent with an orbital period for \gx\ of around
$\sim1.7$\,days.  No other period is simultaneously consistent with
all of the data.  Furthermore, only this period gives a sensible
systemic velocity for the Bowen blend lines.  Bowen radial velocities
reveal a large modulation that must originate from the irradiated
inner face of the companion star.  This implies a projected orbital
velocity of $K_2 = (317 \pm 10)$\,km\,s$^{-1}$ for the preferred
1.7557\,day orbital period (although strictly this is a lower limit).
Hence we derive a mass function of $(5.8\pm0.5)$\,M$_{\odot}$ for the
compact object.  For the shortest acceptable alias of the orbital
period we derive a 95\,\%\ confidence lower limit to the mass function
of 2.0\,M$_{\odot}$.  This makes \gx\ a new addition to the list of
stellar mass black hole candidates supported by dynamical evidence.
The small modulation to the wings of the \HeII\ line suggests a low
mass ratio, $q\la 0.08$, consistent with this interpretation, although
this constraint is less secure.

\acknowledgments

RIH would like to thank Michelle Buxton, Rob Robinson, and Brad Behr
for valuable discussion and criticism.  RIH and PAC acknowledge
support from grant F/00-180/A from the Leverhulme Trust.  RIH is
funded from NASA through Hubble Fellowship grant \#HF-01150.01-A
awarded by STScI, which is operated by the AURA, Inc., for NASA, under
contract NAS 5-26555.  DS acknowledges a PPARC Postdoctoral Fellowship
and a SAO Clay Fellowship.  This work uses observations collected at
ESO in Chile and the AAO in Australia.  We would particularly like to
thank the ESO Director's Office and VLT staff for a generous and
efficiently executed award of Director's Discretionary Time.  This
work has also made use of the NASA ADS Abstract Service.





\begin{table}
\caption{Log of spectroscopic observations of \gx.}
\label{ObsTable}
\begin{tabular}{llclc}
\noalign{\smallskip}
\tableline
\noalign{\smallskip}
Date  & Telescope & UT    & Spectra\tablenotemark{a} & R \\
(2002) &           & range & & (km\,s$^{-1}$) \\
\noalign{\smallskip}
\tableline
\noalign{\smallskip}
Jun  6     & AAT & 10:35--15:11 & 12 & 70 \\
Jun  8     & AAT & 10:46--14:46 &  7 & 70 \\
Jun  9     & NTT & 00:37--09:34 & 24(23) & 60--90 \\
Jun  9     & AAT & 12:01--19:40 & 10 & 70 \\
Jun 10     & NTT & 01:45--05:08 &  7 & 120\\
Jun 10     & AAT & 09:31--10:35 &  3 & 70 \\
Jun 10--11 & NTT & 23:52--00:41 &  2(2) & 60--70 \\
Jun 11     & AAT & 09:59--11:03 &  3 & 70 \\
\noalign{\smallskip}
\tableline
\noalign{\smallskip}
Aug  9--10 & VLT & 23:38--01:31 &  5(5) & 8 \\
Aug 10--11 & VLT & 23:29--01:22 &  5 & 8 \\
Aug 11--12 & VLT & 23:39--01:32 &  5 & 8 \\
Aug 13     & VLT & 00:11--02:04 &  5 & 8 \\
Aug 14     & VLT & 02:13--04:07 &  5 & 8 \\
Aug 14--15 & VLT & 23:56--01:48 &  5(2) & 8 \\
\noalign{\smallskip}
\hline
\end{tabular}
\tablenotetext{a}{Bracketed numbers refer to the number of spectra
  showing convincing sharp \NIII\ components.}
\end{table}

\end{document}